\documentclass[final,5p,times,twocolumn]{elsarticle}

\usepackage{amssymb}
\usepackage{lipsum}
\usepackage{color}
\usepackage{graphicx}
\usepackage{dcolumn}
\usepackage{bm}
\usepackage[utf8]{inputenc}
\usepackage{gensymb}
\usepackage{latexsym}
\usepackage{amsmath}
\usepackage{amsfonts}
\usepackage{nicefrac}
\usepackage{float}
\usepackage{booktabs}
\usepackage{siunitx}
\usepackage{slashed}
\usepackage{hhline}
\usepackage[table]{xcolor}
\usepackage[mathscr,scaled=1.15]{urwchancal}

\usepackage{url}
\usepackage{xspace}
\usepackage{siunitx}
\usepackage{xfrac}
\usepackage{hyperref}
\usepackage[nameinlink]{cleveref}
\usepackage{appendix}

\usepackage{xifthen}
\usepackage{xcolor}
\hypersetup{colorlinks,	linkcolor={green!55!black},	citecolor={green!60!black},	urlcolor={green!55!black}}

\DeclareFontFamily{OT1}{pzc}{}
\DeclareFontShape{OT1}{pzc}{m}{it}%
{<-> s * [1.15] pzcmi7t}{}
\DeclareMathAlphabet{\mathpzc}{OT1}{pzc}{m}{it}

\newcommand{\bracket}[1]{\left( #1 \right)}
\newcommand{\be}{\begin{equation}}
	\newcommand{\bea}{\begin{eqnarray}}
		\newcommand{\ee}{\end{equation}}
	\newcommand{\eea}{\end{eqnarray}}
\newcommand{\sla}{\slash \hspace{-0.22cm}}
\newcommand{\fatg}{{\rm{I}}\!\Gamma}

\biboptions{sort&compress}

\def\1eq#1{Eq.~(\ref{#1})}
\def\ie{{\it i.e.}, }
\def\eg{{\it e.g.}, }
\def\eq#1{Eq.~(\ref{#1})}
\def\2eqs#1#2{Eqs.~(\ref{#1}) and (\ref{#2})}
\def\s#1{{\scriptscriptstyle #1}}

\graphicspath{{./figures/}{./}}

\journal{Physics Letters B}

\begin{document}

\begin{frontmatter}

\title{Heavy-light mesons from a flavour-dependent interaction}

\author[first]{Fei Gao}
\author[second]{Angel S. Miramontes}
\author[second,fourth]{Joannis Papavassiliou}
\author[third,fourth]{Jan M. Pawlowski}

\affiliation[first]{organization={School of Physics, Beijing Institute of Technology},
            city={Beijing},
            postcode={100081}, 
            country={China}}

\affiliation[second]{organization={Department of Theoretical Physics and IFIC, University of Valencia and CSIC},
            city={Valencia},
            postcode={E-46100}, 
            country={Spain}}

\affiliation[third]{organization={
Institut für  Theoretische Physik, Universität Heidelberg},
            city={Philosophenweg 16, Heidelberg},
            postcode={69120}, 
            country={Germany}}

\affiliation[fourth]{organization={ExtreMe Matter Institute EMMI,
GSI},
            city={Planckstrasse 1, Darmstadt},
            postcode={64291}, 
            country={Germany}}

\begin{abstract}
We introduce a new symmetry-preserving framework for the physics of heavy-light mesons, whose key element is the effective  
incorporation of flavour-dependent contributions 
into the corresponding bound-state  
and quark gap equations.
These terms 
originate
from the fully-dressed quark-gluon vertices appearing 
in the kernels of these equations, 
and provide a natural distinction between ``light" and ``heavy" quarks. 
In this approach, only  
the classical
form factor of the quark-gluon vertex is retained, and is 
evaluated in the 
so-called ``symmetric" configuration. 
The standard Slavnov-Taylor identity links  
this form factor to the quark wave-function, allowing for the  
continuous transition from light to heavy quarks 
through the mere variation of 
the current quark mass
in the gap equation. 
The method is used to compute  
the masses and decay constants of specific pseudoscalars and vector heavy-light systems, showing good overall agreement with both experimental data and lattice simulations.
\end{abstract}

\end{frontmatter}




\section{Introduction}
\label{sec:introduction}

The systematic description of heavy-light mesons within the framework of Quantum Chromodynamics (QCD) poses a significant challenge~\cite{Ivanov:1998ms,Bhagwat:2006xi,Gomez-Rocha:2015qga,Fischer:2014cfa,Chen:2019otg,Qin:2019oar,Hilger:2014nma,10.1063/1.3131570,Souchlas:2010zz,Binosi:2018rht,Gelhausen:2013wia,Liu:2015uya,Wang:2015mxa,Tang:2019gvn,Mutuk:2018lki}, mainly due to the substantial mass difference between their constituent quarks. Indeed, 
composed from one heavy  (charm or bottom) and one light (up, down, or strange) valence quark, these mesons are less tractable by standard methods, 
such as the classic  
rainbow-ladder  approximation
and related approaches. 
The experimental determination of the heavy-light meson spectrum, and specifically the discovery of the narrow states $D_s(2317)$~ \cite{BaBar:2003oey, CLEO:2003ggt, Belle:2003guh} and $D_s(2460)$ \cite{CLEO:2003ggt, Belle:2003guh, BaBar:2003cdx}, have attracted significant interest from both the theoretical and experimental communities. These states challenge traditional quark model predictions for heavy-light systems~\cite{DiPierro:2001dwf}, 
in contrast to the better understood $D$ meson states.
Importantly, the heavy-light spectrum will be accessible in the later stages of PANDA~\cite{PANDA:2021ozp}, 
while other experiments, including LHCb~\cite{LHCb:2024nlg,Capriotti:2019huu}, BES III \cite{BESIII:2023wsc} and CMS \cite{CMS:2019uhm}, are conducting precise measurements of cross sections and decay processes involving heavy-light and heavy-heavy mesons. 

The contemporary symmetry-preserving treatment 
of mesonic systems is based 
on the combined analysis 
of Schwinger-Dyson equations (SDEs), and in particular the quark gap equation, and Bethe-Salpeter equations (BSEs)~ \cite{Jain:1993qh,Alkofer:2002bp,
Bender:2002as,
Chang:2009zb,Roberts:1994dr,Savkli:1997kz,Maris:2003vk,Watson:2004kd,Eichmann:2008ae,Qin:2011dd,Fischer:2008wy,Chang:2011ei,Roberts:2011cf,
Bashir:2012fs,Eichmann:2013afa,Heupel:2014ina,
Binosi:2014aea, Sanchis-Alepuz:2015tha,Williams:2015cvx,Bedolla:2015mpa,Sanchis-Alepuz:2015qra,
Li:2016mah,Binosi:2016rxz,
Binosi:2016xxu,Serna:2017nlr,
Miramontes:2019mco,Eichmann:2020oqt,Wallbott:2020jzh,Miramontes:2021xgn,Gutierrez-Guerrero:2021rsx,Yin:2019bxe,Eichmann:2023tjk,Raya:2024ejx}. This set of equations is 
supplied with a propagator-like 
interaction, typically in the form of effective charges or kindred 
quantities. 
It is clear, however, that, 
in order for flavour-effects to be properly taken into account, crucial contributions
stemming from the 
fully-dressed quark-gluon vertices
must be supplemented to the total interaction.

In the present work 
we incorporate such terms at the level of the ``one-gluon exchange” interaction kernel. After the crucial 
skeleton expansion has been duly implemented, this particular kernel contains two fully-dressed 
quark-gluon vertices, whose structure is simplified by retaining only their classical form factor,
evaluated in the {\it symmetric 
configuration}. 
The Slavnov-Taylor identity 
(STI)~\cite{Marciano:1977su,Ball:1980ay} allows us to express 
this form factor 
as the product 
of a universal component
and the flavour-specific 
quark wave-function, whose form is determined from the 
corresponding gap equation. 
The combination of
this universal component 
with the 
gluon propagator gives rise 
to a variant of the standard  
Taylor effective charge~\cite{vonSmekal:1997ohs, Boucaud:2008gn, vonSmekal:2009ae, Blossier:2012ef, Zafeiropoulos:2019flq}, 
which is enhanced by a factor 
of about $1.35$ around the 
$1$ GeV momentum region. 

The subsequent convolution of this 
effective charge with the 
quark wave functions originating from each vertex 
leads to the final 
flavour-dependent interaction strength. Thus,
the relevant flavour content 
is entirely determined from the gap equation,
by adjusting appropriately the values of the current quark masses, 
with no need to resort to additional dynamical equations. Note also that, 
quite importantly,   
the propagator-like nature 
of this interaction 
preserves the crucial 
axial Ward-Takahashi identity (WTI)~
\cite{Roberts:1994dr,Maris:1997hd},
thus enforcing a massless pion in the chiral limit. We emphasize that 
the building blocks of our analysis, 
such as gluon and ghost propagators and the Taylor effective charge, are obtained from lattice QCD~\cite{Aguilar:2021okw,Zafeiropoulos:2019flq}, and 
are in excellent agreement with 
a multitude of functional studies, see, \eg\cite{Mitter:2014wpa, Williams:2015cvx,Cyrol:2016tym,Cyrol:2017ewj,
Huber:2018ned, Gao:2021wun,Ferreira:2023fva}. 
In fact, 
no phenomenological parameter, such as the constituent quark mass or the strength of the vertex dressing, had to be adjusted from experiment.  

The resulting interaction strength is used to estimate the masses and decay constants of several heavy-light systems,
such as $D$, $B$, and $\eta$. 
The calculations are carried out for Euclidean momenta, and the physical values are obtained through an extrapolation algorithm. The results obtained compare very well with experiment and lattice QCD.

\section{Gluons, quarks, and mesons within Landau-gauge QCD}
\label{sec:Mesons}

Our analysis and computations are performed in {\it Landau-gauge} QCD. The main components used in this study 
are introduced below, as items 
({\it i})-({\it v}): \\[-1ex]

({\it i}) The gluon propagator,  \mbox{$\Delta^{ab}_{\mu\nu}(q^2)=-i\delta^{ab}\Delta_{\mu\nu}(q^2)$}, with
\begin{align}
    \Delta_{\mu\nu}(q) = \Delta(q^2)\,P_{\mu\nu}(q)\,, \qquad  \Delta(q^2) = \mathcal{Z}(q^2)/q^2\,,
    \label{eq:gluoprop}
\end{align}
where $P_{\mu\nu}(q) = \delta_{\mu\nu} - 
q_{\mu}q_{\nu}/q^2$ is the transverse projection operator, 
$\Delta(q^2)$ denotes the scalar component of the gluon
propagator, and $\mathcal{Z}(q^2)$ the 
corresponding dressing function. 
At tree-level, $\Delta^{(0)}_{\mu\nu}(q) = P_{\mu\nu}(q)/q^2$,
and so 
$\Delta_{\mu\nu}(q) = \mathcal{Z}(q^2)\,\Delta^{(0)}_{\mu\nu}(q)$.  \\[-1ex]

({\it ii})
The ghost propagator, denoted by 
\mbox{$D^{ab}(q^2) = i\delta^{ab} D(q^2)$}, and the corresponding 
dressing function, $F(q^2)$, 
defined as 
\mbox{$D(q^2) = F(q^2)/q^2$}. \\[-1ex]

({\it iii}) The quark propagator, denoted by \mbox{$S_{\!\!f}^{ab}(p)=i\delta^{ab}S_{\!\!  f}(p)$}, where the index $f$ stands for the quark flavour, 
taking values 
\mbox{$f=u,d,s,c,b$}.
The standard decomposition 
of $S_{\!\!f}^{-1}(p)$
is 
\begin{align} 
\label{eq:qprop}
    S_{\!\!f}^{-1}(p)=i \slashed{p} \, A_{f}(p^2)+B_{f}(p^2)\,,
\end{align}
where $A_{f}(p^2)$ and $B_{f}(p^2)$ are the dressings of the Dirac vector and scalar tensor structure,  respectively. The renormalization-group invariant (RGI)
quark mass function, 
${\mathcal M}_{f}(p^2)$,  
is given by \mbox{${\mathcal M}_{f}(p^2) = B_{f}(p^2)/ A_{f}(p^2)$}.
At tree-level, 
$S_{0,f}^{-1}(p)= i\slashed{p} + m_{f}$\,, 
where $m_{f}$ is the current quark mass for the flavour $f$. Finally, the  
 {\it self-energy}, 
 $\Sigma(p^2)$, is 
 defined as $\Sigma_{f}(p^2) = S_{\!\!f}^{-1}(p) - S_{0,f}^{-1}(p)$.\\[-1ex]

({\it iv}) The fully-dressed quark-gluon vertex 
is written in the form 
\begin{align}
    \fatg_{\! \! f}^{\,a\,\mu}(q,p)=ig\,\frac{\lambda^a}{2} \,
   \Gamma_{\! \!f}^{\,\mu}(q,p)\,, 
    \label{eq:factorG}
\end{align}
where 
$a$ and $\mu$ are the colour and Lorentz indices 
respectively, and $f$ denotes flavour. In addition,   
$\lambda^{a}$, with $a=1,2, ...,8$, are the Gell-Mann matrices, and  
$g$ is the gauge coupling. 
Furthermore, 
$q$ and $p$ denote the incoming gluon and quark momenta respectively; the outgoing anti-quark momentum $r$ is fixed by momentum conservation, 
$r= q+p$. 
At tree-level, the  quark-gluon vertex reduces to $\Gamma^{(0)}_{\mu} = \gamma_\mu$ for all flavours. \\[-1ex]

({\it v})
The {\it Taylor effective charge}, $\alpha_{\s T}(q^2)$, 
defined as ~\cite{vonSmekal:1997ohs, Boucaud:2008gn, vonSmekal:2009ae, Blossier:2012ef, Zafeiropoulos:2019flq} 
\begin{align}
\alpha_{\s T}(q^2) = 
\alpha_s\, {\cal Z}(q^2)\, F^2(q^2) \,,
\label{eq:Taylor} 
\end{align}
where $\alpha_s = g^2/4 \pi$ is the Taylor coupling at the renormalization point $p^2=\mu^2$ with ${\cal Z}(\mu^2)\, F^2(\mu^2)=1$. 
Its modified version \eq{eq:modTaylor} constructed in Sec.\! \ref{sec:interactions} is one of the main building blocks in the present approach.

There are two main dynamical equations that are of central importance for this study, namely the 
quark gap equation that 
controls the evolution of the quark propagator, and the BSE that governs 
the formation of mesonic bound states.

In its renormalized form, the gap 
equation is given by 
\begin{align}
S^{-1}(p)= Z_{2}(i\sla{p}+m_{\s R}) +Z_{1} C_{\!\s F}g^2\int_k\,
\Gamma_{\mu}^{(0)}S(k)\Gamma_{\nu}(q,k)\Delta^{\mu\nu}(q) \,,
\label{eq:gapren}
\end{align}
where $q :=p-k$, and \mbox{$C_{\!\s F}=4/3$} is the Casimir eigenvalue of the fundamental representation, and $m$ is the current quark mass; the flavour index $f$ has been suppressed for simplicity. Furthermore, $\int_{k} := (2\pi)^{-4} \int_{-\infty}^{+\infty} \!\!{\rm d}^4 k$, where the use of a symmetry-preserving regularization scheme is implicitly assumed. Finally, 
$Z_1$ and $Z_2$ are the renormalization constants 
of the quark-gluon vertex and the quark propagator, respectively. Note that \eq{eq:gapren} is complete, in the sense that it contains all possible quantum corrections.

In the meson BSE we will approximate the renormalized four-quark kernel ${\mathcal K}$ by its one-gluon exchange form, $\overline{\mathcal{K}}$,
\begin{align}
\overline{\mathcal{K}} =  g^2 \left( Z_{1} \Gamma_{\mu}^{(0)}\right)  
\Delta^{\mu\nu}(q) \Gamma_{\!\nu}(q,k) \,.
\label{eq:Kbarren}
\end{align}
In \eq{eq:Kbarren} we have paired the classical vertex with the renormalization factor,  $Z_{1} \Gamma_{\mu}^{(0)}$. This building block occurs in all SDEs and BSEs and will be treated within the skeleton expansion, see 
Sec.~\ref{sec:ImprovedScheme}. \eq{eq:Kbarren} is precisely the kernel appearing in  \eq{eq:gapren}. In particular, 
the BSE reads 
\begin{align}
{\cal A}(p,P)= - g^2 \int_{k} \left( Z_{1}
\Gamma_{\!\mu}^{(0)}\right) \,
\tilde{\cal A}(k,P) \,\Gamma_{\!\nu}(q,k) \Delta^{\mu\nu}(q) \,,
\label{eq:BSEoge}
\end{align}
where 
$\tilde{\cal A}(k,P) = S(k_1) {\cal A}(k,P) S(k_2)$, $P$ denotes the total momentum of the meson, $p$ represents the relative momentum between the quark and the anti-quark, $k$ is the loop momentum, 
 with $k_1 =k+P/2$ and $k_2=k-P/2$. As in \eq{eq:Kbarren} we have singled out the universal building block $Z_{1}
\Gamma_{\!\mu}^{(0)}$. In the case of vector 
 mesons, the corresponding 
 BSE is obtained from 
 \eq{eq:BSEoge}
through the substitution 
${\cal A}(p,P) 
\to {\cal A}^{\,\mu}(p,P)$.

\section{Skeleton expansion}
\label{sec:ImprovedScheme}

It is well-known, that an effective charge, such as 
$\alpha_{\s T}(q^2)$, emerges naturally 
inside a one-gluon exchange amplitude if the  
gluon propagator can be combined with momentum-dependent contributions from \textit{both} vertices it is attached to~\cite{Binosi:2009qm,Aguilar:2009nf,Binosi:2014aea,Binosi:2016nme}. 
For the kernel $\overline{\mathcal{K}}$
of \1eq{eq:Kbarren} this entails, that the
momentum-independent component $Z_{1}  \Gamma^{(0)}_{\mu}$ 
must be recast in terms of full vertices and propagators. To that end we resort to the standard skeleton expansion~\cite{Bjorken:1965zz,Roberts:1994dr}. 

There are two main steps employed 
in this procedure. First, the 
standard unrenormalized SDE for $\Gamma$ 
is written such that all 
vertices in its diagrammatic representation 
are fully dressed.
This is accomplished by 
modifying the corresponding multi-particle kernels appearing in the SDE, discarding 
graphs that would cause over-counting, 
such as ladder diagrams.
Schematically,  before renormalization, and
displaying {\it only} 
graphs with quark-gluon vertices, we are led to 
\begin{align}
\Gamma = \Gamma^{(0)} +
\int_k S\, \Gamma^{(0)}\, S M \,=\, \Gamma^{(0)}   + \int_k S \Gamma S 
{\tilde M} \,,
\end{align}
where $M$ is the SDE kernel, while 
${\tilde M}$ is the sum of all fully-dressed 
skeleton graphs,
the first term being  
${\tilde M}_1= g^2\Gamma \Delta \Gamma$. 
By virtue of this rearrangement, 
after renormalization, 
$Z_{1}$ appears only multiplying 
the tree-level term $\Gamma^{(0)}$; thus, 
we get a dressed skeleton relation for $Z_{1} \Gamma^{(0)}$, 
\begin{align}
 Z_{1} \Gamma^{(0)} = \Gamma 
- g^2\!\int_k S \Gamma  
S \Gamma  
\Delta \Gamma + \cdots \,, 
\label{eq:qgsde}
\end{align}
where  
the ellipsis indicates 
({\it a}) one-loop dressed 
graphs with three-gluon and four-gluon vertices, and ({\it b})
higher order diagrams  in the skeleton expansion; they are all comprised exclusively by fully-dressed renormalized propagators and vertices. 

\begin{figure}[t]
\centerline{%
\includegraphics[width=1\columnwidth]{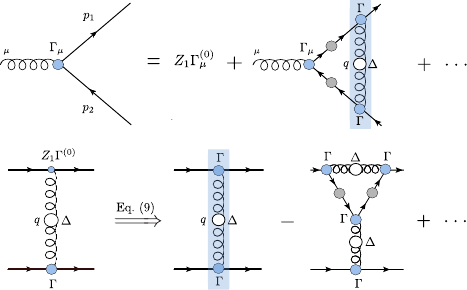}}
\caption{ First row:  the SDE of the quark-gluon vertex in the skeleton expansion;
graphs 
with three- and four-gluon vertices are omitted.
Second row: the 
one-gluon exchange kernel $\bar{\cal K}$ after the use of  \1eq{eq:qgsde}.
The ellipses stand for higher order loop terms.
White and grey circles denote 
fully-dressed gluon and quark propagators, respectively, while 
big (small) blue circles stand for fully-dressed (tree-level) quark-gluon vertices. } 
\label{fig:kernel}     
\end{figure}

The second step of the procedure is to use \eq{eq:qgsde} into \eq{eq:Kbarren}; this generates a form for $\overline{\mathcal{K}}$
which is 
expressed solely in terms of 
fully-dressed propagators and vertices,
to wit
\begin{align}
\overline{\mathcal{K}} 
= g^2
\Gamma_{\mu}(-q,p) \,
\Delta^{\mu\nu}(q)
\, \Gamma_{\nu}(q,k) - \cdots\,,
\label{eq:Kmod}
\end{align}
as shown in Fig.~\ref{fig:kernel}.
When the substitution captured by  Fig.~\ref{fig:kernel} is implemented on the system of 
the gap equation, the meson BSE,
and the SDE of the axial-vector 
vertex, 
one obtains the 
system shown in   Fig.~\ref{fig:BSESDE}, 
where $\overline{\mathcal{K}}$ 
(depicted as a blue box) 
is a common ingredient. 

The ellipses indicate 
higher terms in the 
dressed-loop expansion, which will 
be neglected within the present truncation scheme. 
In fact, 
such additional terms may be interpreted as a systematic RG-improvement; this highlights that the above expansion naturally implements scaling properties, and, in particular, accommodates infrared momentum scaling relevant in the presence of soft (light) modes. We note in passing that the aforementioned  rearrangements may also be derived within the framework 
of $n$-particle irreducible 
effective actions, see \eg~\cite{Baym:1961zz,Cornwall:1973ts, 
Cornwall:1974vz,
Berges:2004pu,
Alkofer:2008tt,Carrington:2010qq, York:2012ib,Williams:2015cvx}, or from the functional renormalization group approach \cite{Pawlowski:2005xe}. 

%

\section{Interaction strengths}
\label{sec:interactions}

\begin{figure}[t!]
\centerline{%
\includegraphics[width=0.45\textwidth]{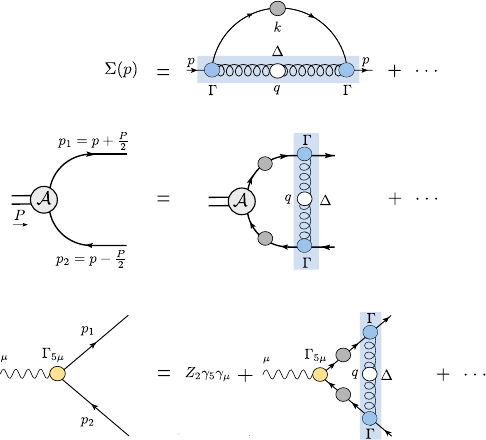}}
\caption{The quark gap equation (first row), 
the meson BSE 
(second row),  
and the SDE of the axial-vector vertex (third row), 
after the implementation of 
\1eq{eq:qgsde}. The yellow circle 
indicates the axial-vector vertex, the wavy line represents a conserved axial-vector current,
and the ellipses denote higher order terms in the skeleton expansion.} 
\label{fig:BSESDE}     
\end{figure}

The accuracy and reliability of the results obtained in the expansion scheme described in Sec.~\ref{sec:ImprovedScheme} hinges on the quantitative determination of the one-gluon exchange scattering kernel \eq{eq:Kmod} that governs all equations, see the blue elongated boxes in Fig.~\ref{fig:BSESDE}. The two quark-gluon 
vertices appearing in \eq{eq:Kmod}
have a double effect 
on the relevant dynamics:
({\it i}) their flavour-independent 
part, identified through a judicious use of the corresponding STI, 
modifies the 
universal part of the interaction, amplifying the 
strength of the standard
Taylor effective charge, $\alpha_{\s T}(q^2)$, and 
({\it ii}) their flavour-dependent parts make the resulting 
interaction sensitive to the type of quarks involved in a given amplitude. 

This analysis, while crucial for the quantitative precision and reliability of the current approach, is rather technical; we therefore highlight the important results: First, the STI allows us to determine the full universal part of the interaction strength as a combination of the Taylor coupling and the universal scalar dressing of the quark-ghost scattering kernel, see \eq{eq:modTaylor}. 
Second, the flavour-dependent part is simply provided by powers of the quark wave functions $A_f(q)$, see \eq{eq:Cfun}. In combination this leads us to the one-gluon scattering kernel of \eq{eq:qq_scat}. 

\begin{figure}[t!]
\centerline{%
\includegraphics[width=0.35\textwidth]{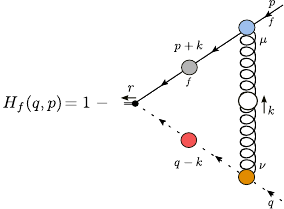}}
\caption{The quark-ghost scattering kernel at 
the one-loop dressed level. The red and orange circles denote the 
fully-dressed ghost propagator and ghost-gluon vertex, respectively.} 
\label{fig:ghost}     
\end{figure}

We proceed with the derivation of these relations. Evidently, a central ingredient in this analysis is the 
STI for the quark-gluon vertex~\cite{Marciano:1977su,Ball:1980ay}, 
\begin{align}
q_\mu \Gamma_{\!f}^\mu(q,p)= F(q^2)\left[ S_{\!f}^{-1}(r)H_{\!f}(q,p) - {\overline H}_{\!f}(-q,r)S_{\!f}^{-1}(p)\right]\,,
\label{eq:STI}
\end{align}
where $H_f$ denotes the composite 
operator known as the ghost-quark scattering kernel (see Fig.~\ref{fig:ghost}), 
corresponding to the quark of 
flavour $f$; 
${\overline H}$ is the 
``conjugate" quantity, obtained from $H_f$ following a set of 
standard rules, see \eg  ~\cite{Davydychev:2000rt,Aguilar:2018csq}. 
The Lorentz decomposition of  
$H_{f}$ is given by~\cite{Davydychev:2000rt} 
\be 
H_{f}(q,p) = X_{0}^{f} 1 + X_1^{f} \slashed{r}
+ X_2^{f} \slashed{p} + X_3^{f}
\sigma_{\mu\nu} r^{\mu} p^{\nu}
\,,
\ee
with the form factors 
$X_{n}^{f} := X_{n}^{f}(q^2,p^2,r^2)$, \,$n=0,1,2,3$. 

We will next keep only the 
classical form factor of the 
quark-gluon vertex, 
namely 
\be
{\Gamma}_{\!f}^{\,\mu}(q,p) 
= {\lambda}_{f}(q,p) \gamma^{\mu} + \cdots\,, 
\label{eq:classic}
\ee
with the ellipsis denoting the 
remaining seven tensorial components. 
As is well-known,   ${\lambda}_{f}(q,p)$
may be expressed 
in terms of the components 
appearing on the r.h.s of 
\eq{eq:STI}; in particular, setting 
$s^2 := r^2 + r\cdot p$  \,(see Eq.~(3.5) in ~\cite{Aguilar:2010cn})
\be
{\lambda}_f = 
\frac{1}{2}F(q^2) \left[A_f(r^2)(X^f_0 -s^2 X^f_3) +B_i(r^2)(X^f_2-X^f_1)\right] + ...
\label{eq:BC}
\ee
where the ellipsis denotes the 
``conjugate" expression, with \mbox{$r \to p$}. 
Given that the $X^f_{1,2,3}$ are numerically subleading~\cite{Aguilar:2016lbe}, an excellent approximation to 
${\lambda}_{f}(q,p)$ is achieved by retaining in \eq{eq:BC}
only the form factor $X^f_{0}$. 

At this point, one notices already the 
emergence of the Taylor effective 
charge, $\alpha_{\s T}(q^2)$, 
defined in \1eq{eq:Taylor}, 
being formed as a combination of 
the gluon propagator and the two factors $F(q^2)$ coming from each vertex. This term is usually referred 
to as ``universal" or ``process-independent", 
since its form does not depend on the details of the specific process studied. 

We next evaluate $\lambda_f(q,p)$ 
in the 
{\it symmetric} configuration, namely  
\mbox{$q^2 =p^2 =r^2$}. 
Denoting the resulting 
function by  $\lambda^{\mathrm{sym} }_f(q^2)$, \eq{eq:BC} turns into  
\begin{align}
{\lambda}^{\mathrm{sym} }_f(q^2)
=  F(q^2) A_{f}(q^2) 
X_0^{f}(q^2) \,, 
\label{eq:gtlsym}
\end{align}
where half of the right hand side comes from the conjugate expression. 
The determination of $X_0^{f}(q^2)$ through the
equation represented by the diagram 
of Fig.~\ref{fig:ghost} 
requires knowledge of the quark 
propagator; the full treatment would therefore entail the coupling of this diagram to the quark gap equation.
However, 
it turns out that $X_0^{f}(q^2)$
contains a flavour-independent part, 
to be denoted by ${\tilde X}_0(q^2)$, which 
emerges when the 
longitudinal part $k_\mu k_\nu$ of the 
gluon propagator $\Delta_{\mu\nu}(k)$
in  Fig.~\ref{fig:ghost}
gets contracted with the quark-gluon vertex [blue circle], 
triggering the STI of \eq{eq:STI}. In particular, 
after setting $H=1$,  \eq{eq:STI} 
furnishes a term $F(k)\,S^{-1}_{\!\!f}(k+p)$,
which removes the internal quark propagator $S_{\!\!f}(k+p)$, yielding 
the $f$-independent contribution ${\tilde X}_0(q^2)$.

With these observations  
it follows straightforwardly that 
\begin{align}
{\tilde X}_0(q^2) = 1 - \frac{1}{2}g^2 N_c  \int_k 
(k \cdot q) \Delta(k^2) D(k^2) B_1(k^2) D(t^2) \,,
\label{eq:X0}
\end{align}
where $t=k+q$ and $N_c$ the number of colours ($N_c=3$).
$B_1(k^2)$ is the tree-level form factor of the ghost-gluon vertex $\Gamma_{\!cg}$
[orange circle in Fig.~\ref{fig:ghost}], 
 renormalized in the Taylor scheme~\cite{Taylor:1971ff,Marciano:1977su,Boucaud:2008gn}
\mbox{($Z_{cg} =1$)}, and 
evaluated 
in the symmetric configuration.

We emphasize that 
the right hand-side of 
\eq{eq:X0} is RGI and hence does not change under renormalization. 
Specifically, 
employing the relations 
$\Delta_{\s R}(q^2) = Z^{-1}_{A} \Delta(q^2)$, 
\mbox{$D_{\!\s R}(q^2) = Z^{-1}_c D(q^2)$}, 
\mbox{$g_{\s R} = Z_A^{1/2} Z_c \,g$}, and $Z_{cg} =1$, we find 
\begin{align} 
g^2 \Delta(k^2) D(k^2) B_1(k^2) D(t^2)= 
g_{\s R}^2 \Delta_{\s R}(k^2) D_{\s R}(k^2) B_1^{\s R}(k^2) D_{\s R}(t^2)\,.
\end{align}
The upshot of the above considerations is to motivate 
the substitution 
\mbox{$X_0^{f}(q^2) \to 
{\tilde X}_0(q^2)$}
inside \eq{eq:gtlsym}. This leads us to 
\begin{align}
{\lambda}^{\mathrm{sym} }_f(q^2)
= F(q^2) {\tilde X}_0(q^2)
A_{f}(q^2) 
\,,
\label{eq:gtlsym2}
\end{align}
where the entire flavour-dependence is contained in 
$A_{f}(q^2)$. 

The kernel $\overline{\mathcal{K}}_{ff'}$ in \eq{eq:Kmod} is constructed from the vertex contributions and the gluon propagator.  
Then, its process-independent part 
corresponds to a modification of the 
Taylor effective charge given in \eq{eq:Taylor}.
Specifically, the inclusion of a factor ${\tilde X}_0(q^2)$ 
from each quark-gluon vertex leads to 
the new effective charge, 
\begin{align}
{\tilde\alpha}_{\s T}(q^2) = 
\alpha_{\s T}(q^2) \,{\tilde X}_0^2(q^2) \,.
\label{eq:modTaylor} 
\end{align}
Note that, since 
both $\alpha_{\s T}(q^2)$
and ${\tilde X}_0(q^2)$
are RGI, so is the 
modified Taylor coupling ${\tilde\alpha}_{\s T}(q^2)$.

\begin{figure*}[t]
\centerline{%
\includegraphics[width=0.7\textwidth]{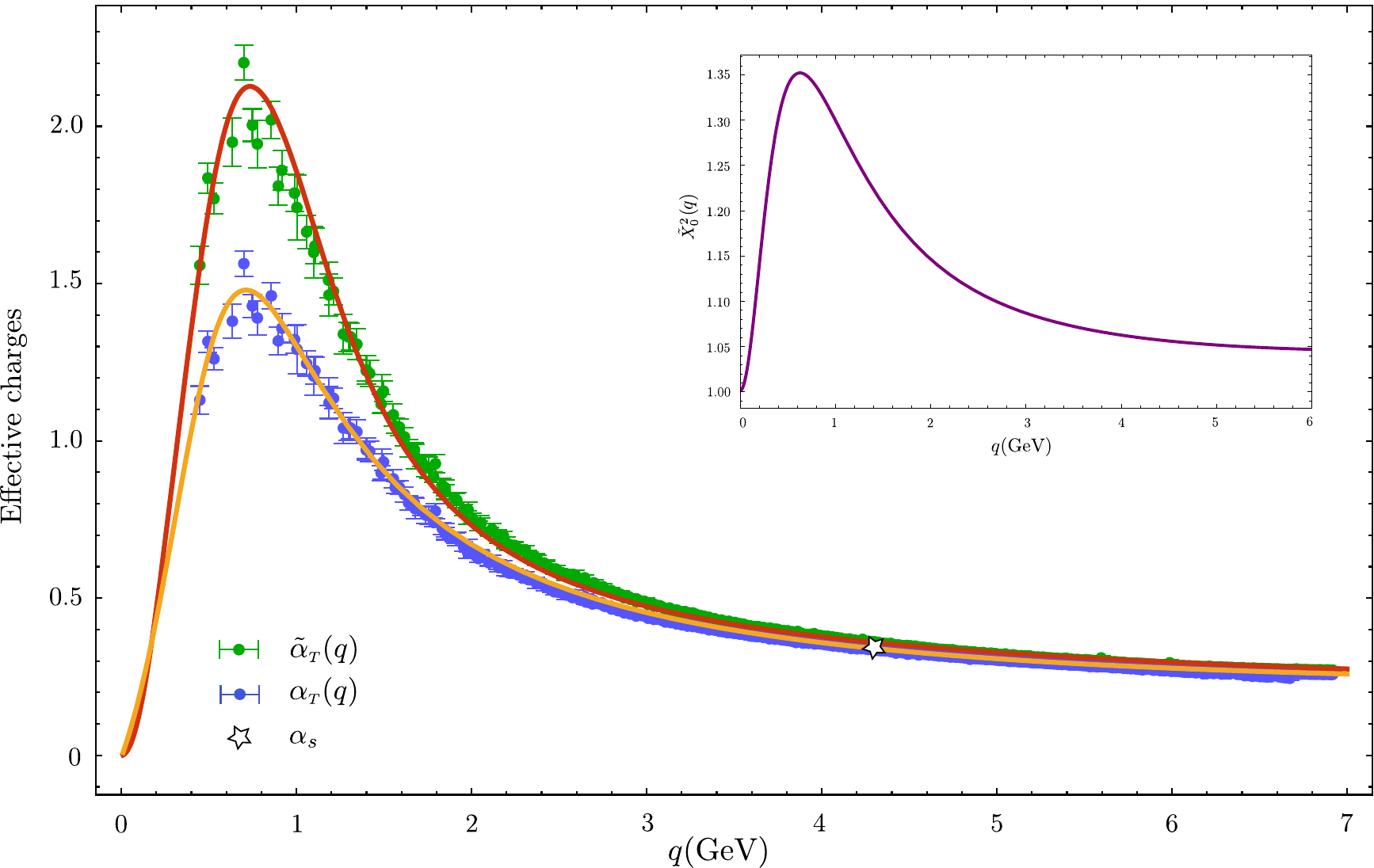}}
\caption{The lattice data of 
\cite{Zafeiropoulos:2019flq}
for the 
Taylor effective charge, $a_{\s T}(q^2)$ (blue data points, with orange continuous line as their fit),
and 
the modified Taylor effective charge, ${\tilde\alpha}_{\s T}(q^2)$,
[green data points], obtained from the 
$a_{\s T}(q^2)$ data through multiplication by ${\tilde X}^2_0(q^2)$, according to \1eq{eq:modTaylor}. 
The red continuous line represents the fit of ${\tilde\alpha}_{\s T}(q^2)$
given by \1eq{eq:dhat_param}. 
The inset 
shows the function ${\tilde X}_0(q^2)$, computed  
by evaluating \1eq{eq:X0}. Finally, 
the ``star" marks the value 
$\alpha_T(4.3 {\rm GeV})$.}
\label{fig:theaTs}     
\end{figure*}

The determination  
of ${\tilde\alpha}_{\s T}(q^2)$
proceeds by providing the 
components $\alpha_{\s T}(q^2)$
and ${\tilde X}_0^2(q^2)$ 
of \eq{eq:modTaylor}. 
For $\alpha_{\s T}(q^2)$ 
we use the lattice data
of \cite{Zafeiropoulos:2019flq},
obtained by combining  
$N=2+1+1$ 
gluon and ghost propagators 
according to \eq{eq:Taylor},
see Fig.~\ref{fig:theaTs}.
The ghost-quark component 
${\tilde X}_0(q^2)$ is 
obtained from \eq{eq:X0}, 
where the 
lattice results 
of \cite{Zafeiropoulos:2019flq}
for $\Delta(k^2)$ and $D(k^2)$
are employed. 
$B_1(k^2)$ is computed from Eq.(4.5) of  \cite{Aguilar:2021okw}, 
after the appropriate inclusion of quark effects.
Finally, the coupling $\alpha_s$ used in \eq{eq:X0}
is the central value 
of $\alpha_T(q)$ 
at $q=4.3$ GeV, 
extracted from the orange curve 
in Fig.~\ref{fig:theaTs} (denoted by a ``star"), 
namely \mbox{$\alpha_T(4.3 {\rm GeV}) = 0.354 \pm 0.007$}. We stress that the 
above determination of 
${\tilde\alpha}_{\s T}(q^2)$ 
contains {\it no} adjustable parameters.

For computational convenience,  the resulting curve of ${\tilde\alpha}_{\s T}(q^2)$, 
shown in 
Fig.~\ref{fig:theaTs} (red line), 
is fitted using 
the functional form, 
\be
{\tilde\alpha}_{\s T}(q^2)
= \frac{a_0 q^2 + a_1 q^4 \ln\left(1 + \frac{\Lambda_0^2}{q^2}\right) + a_2 q^4}{1 + a_3 q^2 + a_4 q^4 + a_5 q^6}
+ \frac{4 \pi q^6}{\beta_0 \left(\Lambda_0^6 + q^6 \ln \frac{q^2}{\Lambda_T^2}\right)},
\label{eq:dhat_param}
\ee    
with the values
$a_0 = 10.35$ GeV$^{-2}$, $a_1=23.69$ GeV$^{-4}$, $a_2=27.94$ GeV$^{-4}$, $a_3=10.72$ GeV$^{-2}$, $a_4=-2.5$ GeV$^{-4}$ and $a_5=29.02$ GeV$^{-6}$. Additionally, $\Lambda_T$=0.5 GeV, $\Lambda_0=1$ GeV and $\beta_0=11-2n_f/3$ ($n_f=4$).

Once ${\tilde\alpha}_{\s T}(q^2)$
has been determined,
the incorporation of  $A_{f}(q^2)$ and
$A_{f'}(q^2)$ from each vertex gives rise to  
the full {\it flavour-dependent} interaction, 
\begin{align}
{\mathcal I}_{\!ff'}(q^2)
 = 
{\tilde\alpha}_{\s T}(q^2) A_{f}(q^2) 
A_{f'}(q^2) \,.
\label{eq:Cfun}
\end{align}
We emphasize that, 
in contradistinction to 
$\alpha_{\s T}(q^2)$
and ${\tilde\alpha}_{\s T}(q^2)$, the 
${\mathcal I}_{\!ff' }(q^2)$
is {\it not} RG-invariant;
specifically, we have that 
\be
{\mathcal I}_{\!ff'}^{R}(q^2) = 
Z_{2,\,f}^{-1} \,Z_{2,\,f'}^{-1}\,{\mathcal I}_{\!ff'}(q^2) \,.
\label{eq:Cren}
\ee
Then, the kernel 
$\overline{\mathcal{K}}_{ff'}$, 
given in \1eq{eq:Kmod}, 
may be expressed 
in terms of the interaction ${\mathcal I}_{\!ff'}(q^2)$
according to 
\be
\overline{\mathcal{K}}_{ff'} = 
4\pi \gamma^{\mu} 
{\mathcal I}_{\!ff'}(q^2)
\gamma^{\nu} \Delta^{(0)}_{\mu\nu}(q)\,.
\label{eq:qq_scat}
\ee
We may now cast both 
\eq{eq:gapren} and \eq{eq:BSEoge} 
in terms of ${\mathcal I}_{\!ff'}(q^2)$,
namely
\begin{equation}
S_{\!f}^{-1}(p)= Z_{2,\,f}(i\slashed{p}+m_f) + 4 \pi C_{\!\s F} \!\!\int_k\,
\gamma^{\mu} S_{\!f}(k)\gamma^{\nu} 
\Delta^{(0)}_{\mu\nu}(q) 
\,
{\mathcal I}_{\!ff }(q^2)
\label{eq:gapI}
\end{equation}
and
\begin{align}
{\cal A}_{ff'}(p,P)= 
-4 \pi \int_{k} \gamma^{\mu}
{\tilde{\cal A}}_{ff' }(k,P) 
\gamma^{\mu} \Delta^{(0)}_{\mu\nu}(q)
{\mathcal I}_{\!ff' }(q^2)
 \,,
\label{eq:BSEI}
\end{align}
while, for the BSE of the vector mesons, we simply set 
\mbox{${\cal A}(p,P) 
\to {\cal A}_{\mu}(p,P)$}
at the level of \eq{eq:BSEI}.
We note that, 
since \mbox{$\tilde{\cal A}(k,P)=S(k_1){\cal A}(k,P)S(k_2)$}, 
the full kernel 
of \eq{eq:BSEI} is RGI by virtue of 
\eq{eq:Cren}, 
as it should.

We end this section by 
stressing 
that the procedure 
leading to
\eq{eq:gapI} and \eq{eq:BSEI} 
is {\it symmetry preserving}, in the sense that the WTI 
satisfied by the axial-vector vertex, 
$\Gamma_{\!\!5\mu}^{f}(p_1, p_2)$, in the chiral limit \cite{Qin:2014vya},
\ie
\begin{align}
P^{\mu} \Gamma_{5\mu}^{\,f}(p_1, p_2) =  S^{-1}_{\!\!f}(p_1)  \gamma_5 +  \gamma_5 S^{-1}_{\!\!f}(p_2) \,,
\label{eq:AWTI} 
\end{align}
is exactly fulfilled, with 
$f$ a flavour index. 
Indeed, in the 
one-loop dressed approximation, 
the SDE of this vertex
involves precisely the 
kernel ${\mathcal I}_{\!f f}(q^2)$, 
namely 
\begin{align}
\Gamma^{\,f}_{\!\!5\mu}(p_1, p_2) =  \gamma_5 \gamma_{\mu} - 
4 \pi\int_k 
\gamma^{\alpha}
{\tilde\Gamma}^{\,f}_{5 \mu}(k_1,k_2) \, 
\gamma^{\beta} \Delta^{(0)}_{\alpha\beta}(q)
{\mathcal I}_{\!ff}(q^2)\,, 
\label{eq:axvert}
\end{align}
with  ${\tilde\Gamma}_{5\mu}^{\,f}(k_1,k_2) = S_{\!f}(k_1) \Gamma_{\!\!5\mu}^{\,f}(k_1,k_2)S_{\!f}(k_2)$. 
Consequently, 
the contraction 
of both sides of 
\eq{eq:axvert} by 
$P^{\mu}$, 
and the use of 
\eq{eq:AWTI}
under the integral sign,
returns precisely 
the difference of 
two quark 
self-energies
[{\it viz.} \eq{eq:gapI}], 
exactly as happens 
in the case of the 
rainbow-ladder 
approximation. 
Clearly, this
demonstration goes through because 
$({\it i})$
the kernel 
${\mathcal I}_{\!ff }(q^2)$ is 
{\it common} to both 
\eq{eq:gapI}
and \eq{eq:axvert}, 
and $({\it ii})$ due to the 
choice of the {\it symmetric limit}, 
its momentum dependence is the same 
as that of the gluon propagator. The preservation of 
\eq{eq:AWTI} guarantees 
the vanishing of the pion mass in the chiral 
limit~\cite{Roberts:1994dr,Maris:1997hd}, which is  
confirmed explicitly in the 
numerical analysis below. 

We emphasize, however,  
that, in the presence of 
non-vanishing current 
quark masses, $m_f$ and $m_{f'}$, 
the generalized WTI identity  
(see Eq.(2.5) in \cite{Roberts:2015lja})
is not satisfied by the interaction ${\mathcal I}_{\!ff'}(q^2)$
in \1eq{eq:Cfun}; for a symmetry-preserving treatment with a vertex solely depending on the gluon momentum see e.g.~\cite{Xu:2022kng}. 
The violations are proportional to  
$\left|m_f-m_{f'}\right|$, which can  be easily established by repeating the  
steps listed below \1eq{eq:axvert}, and noting that the term $A_f (q^2) A_{f'}(q^2)$ may not be assigned to the self-energy of either $S_{\!\!f}(p_1)$ or $S_{\!\!f'}(p_2)$. A consequence of the above structure are deviations from the 
Gell-Mann--Oakes--Renner relation~\cite{Gell-Mann:1968hlm}, which is part of the systematic error of the current approximation. 

\begin{figure*}[t]
\centerline{%
\includegraphics[width=0.8\textwidth]{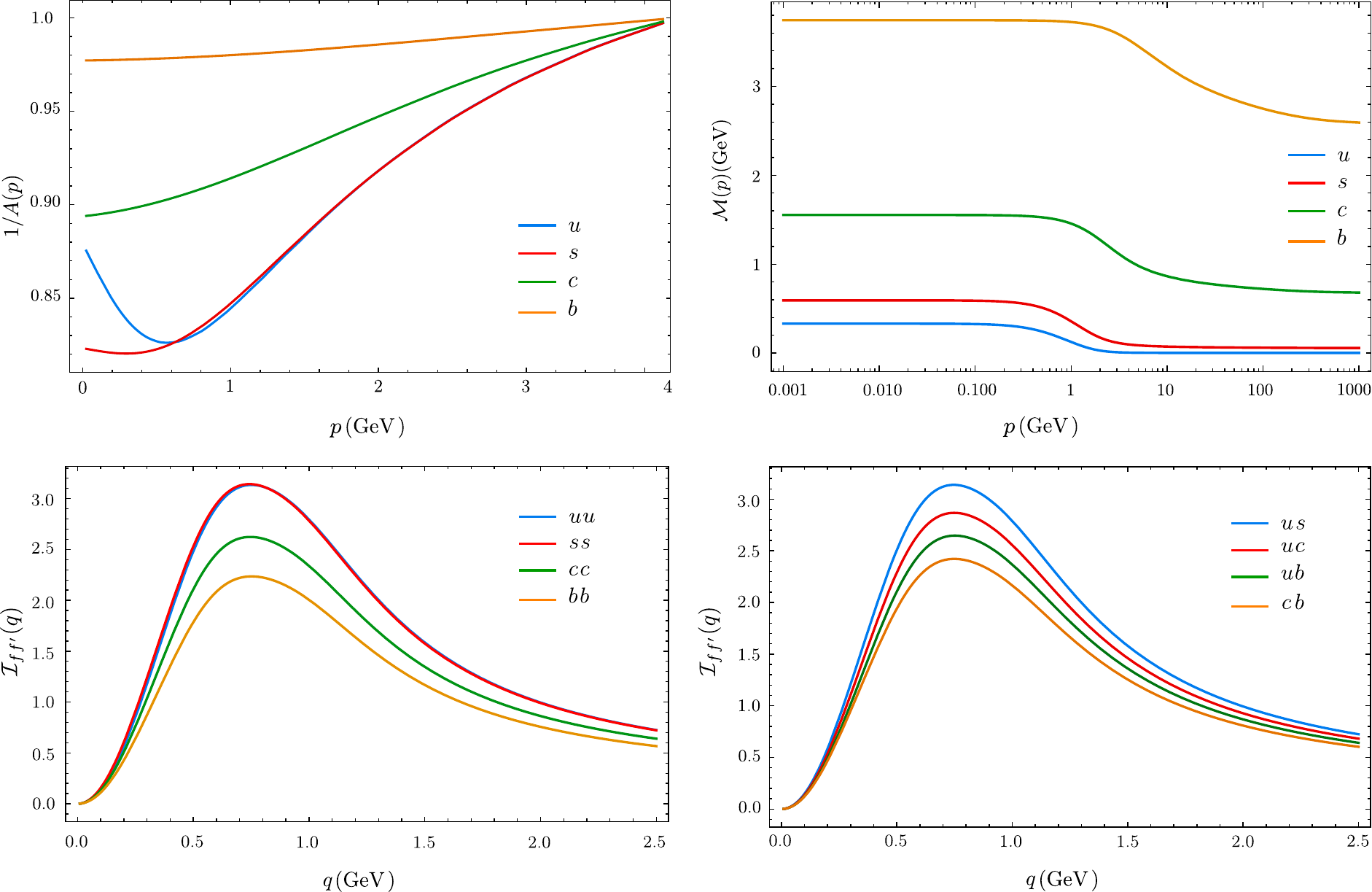}}
\caption{The inverses of the quark wave functions (upper left); 
the constituent quark masses ${\cal M}(p^2)$ (upper right);
direct comparison of the  four interaction strengths ${\cal I }_{f f}(q^2)$ 
(lower left);
comparison between 
${\mathcal I}_{\!uu}(q^2)$, 
${\mathcal I}_{\!cc}(q^2)$, 
and 
${\mathcal I}_{\!cu}(q^2)$
(lower right).
} 
\label{fig:I_f1f2}     
\end{figure*}
%

\section{Results}
\label{sec:Results} 

In this section we 
use the two 
main dynamical equations, \eq{eq:gapI} and \eq{eq:BSEI}, 
to evaluate 
the ground-state masses of both light and heavy-light mesons, along with the corresponding decay constants,
given by \cite{Maris:1997hd,Maris:1999nt},
\begin{align}
    f_{ \text{ps}} m^2_{\text{ps}} =&\, \sqrt{N_c} Z_2 \text{Tr} \int_k \gamma^5 \slashed{P} S(k_1) \mathcal{A}\bracket{k,P}S(k_2) \,,
    \nonumber \\
    f_{\text{v}} m_{\text{v}} =&\, \frac{\sqrt{N_c}}{3} Z_2 \text{Tr} \int_k  \gamma_\mu S(k_1) \mathcal{A}_{\mu} \bracket{k,P}S(k_2)\,,
\end{align}
Before entering into the 
technicalities of 
the numerical treatment, 
we briefly comment on the 
renormalization procedure.  
After taking the appropriate traces, 
\eq{eq:gapI} yields two coupled 
integral equations 
for $A_{f}(p^2)$ and 
$B_{f}(p^2)$. These quantities depend on the renormalization constants 
$Z_2$ and $Z_m$, 
to be fixed by 
employing a specific renormalization scheme.
In particular, in the standard 
momentum subtraction 
(MOM) scheme with the renormalization point $p^2 = \mu^2$, one imposes 
 the conditions 
\begin{align}
A_{f}(\mu^2) =1 \,,
\qquad
B_{f}(\mu^2) = m_f\,, 
\label{eq:mom}
\end{align}
which, in turn, provide 
expressions for $Z_2$ and $Z_m$ in terms of integrals evaluated at 
$q^2 = \mu^2$. 
The value of the $\mu$
chosen for the actual 
calculations is
$\mu =4.3$ GeV.
Note that, while  
${\cal M}_i(p^2)$ 
is RGI, the 
$\mu$-dependence of 
$A_{f}(q^2)$ is 
transmitted to the 
interaction strength 
${\mathcal I}_{\!ff'}(q^2)$, which also 
depends on $\mu$. 

The central quantity appearing in 
both \eq{eq:gapI} and \eq{eq:BSEI} 
is the flavour-dependent interaction  
${\mathcal I}_{\!ff'}(q^2)$ 
of \eq{eq:Cfun}. Importantly, 
the components that carry the 
dependence on the flavour, namely 
$A_{f}(q^2)$ and $A_{f'}(q^2)$, are dynamically determined by solving \eq{eq:BSEI}.

The numerical procedure
adopted 
is summarized below; for a detailed description of the numerical techniques and algorithms, the reader is referred to~\cite{Eichmann:2016yit,Sanchis-Alepuz:2017jjd}:\\[-2ex]

({\it a}) In Euclidean space-time, the total momentum $P$ 
of a bound state of mass $M$  
is parametrized as \mbox{$P = (0,0,0,i M)$}, where \mbox{$P^2=-M^2$} 
Then, the arguments of the 
quark propagators $S(k_{1,2})$ 
appearing in the 
BS equation satisfy 
\mbox{$k^2_{1,2}=k^2+P^2/4 \pm k\cdot P$}, corresponding to parabolas in the complex plane
formed by $k_1$ and $k_2$. 
The analytic continuation of the gap equation to the complex plane is achieved using the Cauchy
interpolation method \cite{Sanchis-Alepuz:2017jjd,Fischer:2005en,Krassnigg:2008bob}. 
The range of its applicability 
is restricted by the appearance
of complex conjugate poles in $S(k_{1,2})$; their 
exact position, and the 
maximum mass 
that can be extracted, depend on the values of the current quark masses, $m_i$.\\[-2ex]

\begin{table*}[t!]
    \centering
    \sisetup{table-format=1.6(2)}
    \renewcommand{\arraystretch}{1.4}
    \rowcolors{6}{gray!25}{white}
    \setlength{\tabcolsep}{4.5pt} 
    \arrayrulecolor{black!40} 
    \begin{tabular}{|l|*{16}{S[table-format=1.3]|}} 
        \hline
        \rowcolor{gray!10}
        {mass (GeV)} & {$\pi$} & {$K$} & {$\rho$} & {$D$} & {$D^*$} & {$D_s$} & {$D_s^*$} & {$B$}  & {$B^*$} & {$B_s$} & {$B_s^*$} & {$B_c$} & {$B_c^*$} & {$\eta_c$} & {$J/\Psi$} &{$\eta_b$}\\
        \hline
        {This work} & {0.139$^{\dagger}$} & {0.495$^{\dagger}$} & {0.735} & {1.93} & {2.06}  & {2.03} & {2.17} & {5.33}  & {5.37} & {5.45} & {5.48} & {6.36} & {6.40} & {2.97$^{\dagger}$} &{3.13} & {9.5$^{\dagger}$}\\
        \hline
        {Experiment \cite{ParticleDataGroup:2022pth}} & {0.139} & {0.494} & {0.775} & {1.87} & {2.01}  & {1.97} & {2.11} & {5.28}  & {5.33} & {5.37} & {5.42} & {6.28} & {} & {2.98} & {3.10}& {9.4}\\
        \hline
        {\% deviation} & {} & {} & {5.1} & {3.2} & {2.5}  & {3.0} & {2.8} & {0.9}  & {0.7} & {1.5} & {1.1} & {1.3} & {} & {} & {1.0} & {}\\
        \hhline{|=|=|=|=|=|=|=|=|=|=|=|=|=|=|=|=|=|}{Lattice~\cite{Lubicz:2017asp,Dowdall:2012ab,Cichy:2016bci,Mathur:2018epb,Donald:2012ga}} & {} & {} & {} & {1.86} & {2.01}  & {1.97} & {2.11} & {5.28}  & {5.32} & {5.37} & {5.42} & {6.26} & {6.33} & {} &{3.12}& {}\\
        \hline
    \end{tabular}
    \caption{Meson masses, all results are presented in GeV. The free parameters have been fixed to reproduce the values denoted with a $\dagger$.}
    \label{tab:masses}
\end{table*}
\begin{table*}[t!]
    \centering
    \sisetup{table-format=1.4(2)}
    \renewcommand{\arraystretch}{1.45}
    \setlength{\tabcolsep}{2.05pt} 
    \arrayrulecolor{black!40} 
    \begin{tabular}{|l|*{16}{S[table-format=1.3]|}} 
       \hline
        \rowcolor{gray!10}
        {$f$ (GeV)} & {$\pi$} & {$K$} & {$\rho$} & {$D$} & {$D^*$} & {$D_s$} & {$D_s^*$} & {$B$}  & {$B^*$} & {$B_s$} & {$B_s^*$} & {$B_c$} & {$B_c^*$} & {$\eta_c$} & {$J/\Psi$} &{$\eta_b$}\\
           \hline
        {This work} & {0.130$^{\dagger}$} & {0.158} & {0.205}  & {0.201}  & {0.218} & {0.249} & {0.270}  & {0.201} & {0.197} & {0.235} & {0.231} & {0.505} & {0.492} & {0.355} &{0.433}& {0.762}\\
           \hline
            {Experiment \cite{ParticleDataGroup:2022pth}} & {0.130} & {0.155} & {0.212} & {0.196} & {}  & {0.236} & {} & {0.211}  & {} & {} & {} & {} & {} & {0.340} & {0.415}&{}\\
       \hline
        {\% deviation} & {} & {1.9} & {3.3} & {2.6} & {}  & {5.5} & {} & {4.7}  & {} & {} & {} & {} & {} & {4.4} &{4.0}& {}\\
       \hhline{|=|=|=|=|=|=|=|=|=|=|=|=|=|=|=|=|=|}
        {ETM~\cite{Lubicz:2017asp}} & {} & {} & {} & {0.208} & {0.223}  & {0.247} & {0.268} & {0.193}  & {0.186} & {0.229} & {0.223} & {} & {} & {} & {}&{}\\  
         \hline
      {MILC18~\cite{Bazavov:2017lyh}} & {} & {} & {} & {0.213} & {}  & {0.245} & {} & {0.189}  & {} & {0.231} & {} & {} & {} & {} & {}&{}\\ 
         \hline
{HPQCD~\cite{Davies:2010ip,Hughes:2017spc,McNeile:2012qf,Donald:2012ga}} & {0.132} & {0.157} & {} & {0.207} & {}  & {0.241} & {} & {0.189}  & {} & {0.231} & {} & {0.427} & {0.422} & {0.395} &{0.404}& {}\\ 
       \hline
         {RBC/UKQCD~\cite{Boyle:2017jwu,Christ:2014uea}} & {} & {} & {} & {0.209} & {}  & {0.246} & {} & {0.196}  & {} & {0.235} & {} & {} & {} &{}& {} & {}\\   \hline
    \end{tabular}
    \caption{Decay constants, all results are presented in GeV. The free parameters have been fixed to reproduce the values denoted with a $\dagger$. In our normalization, $f_{\pi} = 0.130$ GeV.}
    \label{tab:decay_constants} 
\end{table*}

({\it b}) The procedure adopted for fixing the current quark masses, as well as the pion and 
$\eta$-meson masses, is as follows. 
First, we compute the quark propagator for a given current quark mass on the complex plane, using the Cauchy interpolation method. 
Since the pion mass lies within the computable region of the propagator, the $u$
and $d$ masses, assumed to be equal, can be easily adjusted to reproduce the experimental value. 
In the case of the $\eta$ meson, its mass lies 
in a region where 
the quark propagator cannot be directly computed, due to the presence of complex poles. Therefore, for that region, 
an Ansatz containing a 
pair of 
complex conjugate poles is used,
namely \cite{Rojas:2014aka,Mojica:2017tvh}, 
\begin{align}\nonumber 
S(p) =&\, -i \slashed{p} \sigma_v(p^2) + \sigma_s(p^2) \, , \\[1ex]\nonumber
\sigma_v (p^2) =&\, \sum_{i}^n \left[\frac{\alpha_i}{p^2 + m_i} + \frac{\alpha_i^\ast}{p^2 + m_i^\ast}\right]  \, , \\[1ex]
\sigma_s (p^2) =&\, \sum_{i}^n \left[\frac{\beta_i}{p^2 + m_i} + \frac{\beta_i^\ast}{p^2 + m_i^\ast}\right] \,,
\label{eq:poles_ansatz}
\end{align}
where the parameters $m_i$, $\alpha_i$, and $\beta_i$ are derived by fitting the respective quark SDE solution along the parabolic trajectory in the complex plane. If the calculated bound state mass for the chosen current quark mass fails to reproduce the $\eta$ meson mass, the entire procedure must be iterated, for a new set of quark masses.  The current 
quark masses obtained through this procedure are $m_{u/d} = 0.005$ GeV, $m_s = 0.094$ GeV, $m_c = 1.1$ GeV, and $m_b = 3.5$ GeV, at  $\mu = 4.3$ GeV.\\[-2ex]

({\it c}) The BS equation is solved numerically by reformulating it into an eigenvalue problem. Physical solutions correspond to the mass-shell points $P^2_n = -M^2_n$, where $M^2_0$ denotes the ground state mass of the meson, and $M^2_n$ ($n \geq 1$) represents the $n^{\rm th}$ radial excitation. Furthermore, the calculation of meson BS amplitudes is simplified by expanding them into Chebyshev polynomials of the second kind, thus facilitating the factorization of the angular dependence. For the calculation of all meson BS amplitudes (both pseudoscalar and vector particles), we have employed a total of 10 such polynomials. We remark that, 
in the case of the heavy-light mesons, 
the significant differences in quark masses require high precision to ensure convergence. In particular, we have employed a total of 160 integration points for the relative momentum $k$ and 48 points for the angular variables. Additionally, the heavy-light and heavy-heavy meson masses have been computed employing the quark parametrization from \eq{eq:poles_ansatz}.\\[-2ex]

The results of this analysis may be summarized as follows: \\[-2ex]

({\it i}) 
The inverse quark wave functions 
$A^{-1}_{f}(p^2)$ for 
$f=u,s,c,b$ 
are shown 
in the left upper panel 
of Fig.~\ref{fig:I_f1f2}; 
the case $f=d$ coincides with that of $u$, and is omitted.  
We note that 
$A^{-1}_{u}(p^2)$ displays the 
typical minimum, 
which is hardly visible in $A^{-1}_{s}(p^2)$, and absent in 
$A^{-1}_{c}(p^2)$ and $A^{-1}_{b}(p^2)$; again, 
$\mu= 4.3$ GeV. \\[-2ex]

({\it ii}) 
The dynamically generated quark masses 
${\cal M}_i(p^2)$ 
are shown in the upper right panel 
of Fig.~\ref{fig:I_f1f2}. 
The corresponding values at the 
origin are ${\cal M}_u(0) = 0.34$ GeV, 
${\cal M}_s(0) = 0.59$ GeV,
${\cal M}_c(0) = 1.55$ GeV, and 
${\cal M}_b(0) = 3.78$ GeV.\\[-2ex]

({\it iii}) 
The diagonal elements, $f=f'$,  
of the 
flavour-dependent interaction strength 
${\mathcal I}_{\!ff'}(q^2)$, 
are shown in the 
lower left panel of 
Fig.~\ref{fig:I_f1f2}, for $f=u,s,c,b$, 
and  with
$\mu = 4.3$ GeV.  
Note that ${\mathcal I}_{\!uu}(q^2) \approx 
{\mathcal I}_{\!ss}(q^2)$. 
As expected, 
the intensity of the interaction decreases when heavier 
quarks are considered, \ie 
\mbox{${\mathcal I}_{\!bb}(q^2) < 
{\mathcal I}_{\!cc}(q^2) <
{\mathcal I}_{\!ss}(q^2)$} 
for all momenta except the origin,
where ${\mathcal I}_{\!ff'}(0)=0$
for all cases. 
Furthermore, in the 
lower right panel of 
Fig.~\ref{fig:I_f1f2}
we show some of the 
non-diagonal 
elements. Comparison 
with the curves in the lower left panel 
reveals that, as expected,   
${\mathcal I}_{\!cc}(q^2) < 
{\mathcal I}_{\!cu}(q^2) <
{\mathcal I}_{\!uu}(q^2)$, with analogous
relations for all other 
combinations considered.\\[-2ex]

({\it iv})
The masses and decay constants 
of 16 states are shown in 
Tables \ref{tab:masses} and 
\ref{tab:decay_constants}, respectively,
where they are compared with the 
corresponding experimental and lattice 
values, whenever available.
We find excellent agreement, 
reflected in the small percentage 
deviations
reported, which do not 
exceed the few percent level.
The most notable exceptions 
are the decay constants 
of $B_c$, $B_c^*$ and $\eta_c$, 
whose relative deviations from the lattice results are 18$\%$, 17$\%$, and  
10$\%$, respectively. \\[-2ex]

({\it v})
Finally, we have computed the masses of the first radial excitations for the pion and kaon, obtaining $m_{\pi_1} = 1.24$GeV and $m_{K_1}=1.30$GeV, 
compared
with the experimental values of $m^{\s {\rm exp}}_{\pi_1} = 1.30$GeV and $m^{\s {\rm exp}}_{K_1}=1.46$GeV \cite{ParticleDataGroup:2022pth}. 
Note that the above mass hierarchy  
is typically inverted within 
the rainbow-ladder approximation, 
requiring the inclusion of the anomalous 
chromomagnetic moment for its restoration \cite{Qin:2020jig,Xu:2022kng}.

\section{Conclusions}
\label{sec:conclusions}

In this work we have introduced a novel expansion scheme for functional bound state equations on the basis of a four-quark kernel solely derived from QCD correlation functions: The two main building blocks of this kernel are a universal flavour-independent effective charge, see \eq{eq:modTaylor}, and flavour-dependent contributions in terms of the quark dressings, leading to the final kernel \eq{eq:Cfun}.   
These two central quantities can either be obtained self-consistently from solutions of functional approaches to QCD, or from lattice simulations. 

The quantitative precision of this scheme has been established within the computation of masses and decay constants of heavy-light mesons, see Tables \ref{tab:masses} and 
\ref{tab:decay_constants}. We envisage its application to a broad range of applications, and in particular 
on electromagnetic meson form factors and parton distribution amplitudes. Moreover, while improving the expansion scheme beyond the present single gluon exchange approximation in a symmetry-preserving way proves challenging, we hope to report on this in the near future.

\section*{Acknowledgements}

We thank G.~Eichmann, M.N.~Ferreira and 
J.~Rodriguez-Quintero
for discussions. The work of A.S.M.~and 
J.P.~is funded by the Spanish MICINN grants PID2020-113334GB-I00 and
PID2023-151418NB-I00,  
the Generalitat Valenciana grant CIPROM/2022/66,
and CEX2023-001292-S by MCIU/AEI.
J.P.~is supported 
in part by the EMMI visiting grant of 
the ExtreMe Matter Institute EMMI
at the GSI,
Darmstadt, Germany.
F.G.~is supported by the National  Science Foundation of China under Grants  No. 12305134. J.M.P.~is  funded by  the Deutsche Forschungsgemeinschaft (DFG, German Research Foundation) under Germany’s Excellence Strategy EXC 2181/1 - 390900948 (the Heidelberg STRUCTURES Excellence Cluster) and the Collaborative Research Centre SFB 1225 - 273811115 (ISOQUANT).

\bibliographystyle{elsarticle-num_back}
\bibliography{HeavyLight}

\begin{thebibliography}{100}
\expandafter\ifx\csname url\endcsname\relax
  \def\url#1{\texttt{#1}}\fi
\expandafter\ifx\csname urlprefix\endcsname\relax\def\urlprefix{URL }\fi
\expandafter\ifx\csname href\endcsname\relax
  \def\href#1#2{#2} \def\path#1{#1}\fi

\bibitem{Ivanov:1998ms}
M.~A. Ivanov, Y.~L. Kalinovsky, C.~D. Roberts, Phys. Rev. D 60 (1999) 034018.

\bibitem{Bhagwat:2006xi}
M.~S. Bhagwat, A.~Krassnigg, P.~Maris, C.~D. Roberts, Eur. Phys. J. A 31 (2007) 630--637.

\bibitem{Gomez-Rocha:2015qga}
M.~Gomez-Rocha, T.~Hilger, A.~Krassnigg, Phys. Rev. D 92~(5) (2015) 054030.

\bibitem{Fischer:2014cfa}
C.~S. Fischer, S.~Kubrak, R.~Williams, Eur. Phys. J. A 51 (2015) 10.

\bibitem{Chen:2019otg}
M.~Chen, L.~Chang, Chin. Phys. C 43~(11) (2019) 114103.

\bibitem{Qin:2019oar}
P.~Qin, S.-x. Qin, Y.-x. Liu, Phys. Rev. D 101~(11) (2020) 114014.

\bibitem{Hilger:2014nma}
T.~Hilger, C.~Popovici, M.~Gomez-Rocha, A.~Krassnigg, Phys. Rev. D 91~(3) (2015) 034013.

\bibitem{10.1063/1.3131570}
T.~Nguyen, N.~A. Souchlas, P.~C. Tandy, AIP Conference Proceedings 1116~(1) (2009) 327--333.

\bibitem{Souchlas:2010zz}
N.~Souchlas, Phys. Rev. D 81 (2010) 114019.

\bibitem{Binosi:2018rht}
D.~Binosi, L.~Chang, M.~Ding, F.~Gao, J.~Papavassiliou, C.~D. Roberts, Phys. Lett. B 790 (2019) 257--262.

\bibitem{Gelhausen:2013wia}
P.~Gelhausen, A.~Khodjamirian, A.~A. Pivovarov, D.~Rosenthal, Phys. Rev. D 88 (2013) 014015, [Erratum: Phys.Rev.D 89, 099901 (2014), Erratum: Phys.Rev.D 91, 099901 (2015)].

\bibitem{Liu:2015uya}
J.-B. Liu, M.-Z. Yang, Chin. Phys. C 40~(7) (2016) 073101.

\bibitem{Wang:2015mxa}
Z.-G. Wang, Eur. Phys. J. C 75 (2015) 427.

\bibitem{Tang:2019gvn}
S.~Tang, Y.~Li, P.~Maris, J.~P. Vary, Eur. Phys. J. C 80~(6) (2020) 522.

\bibitem{Mutuk:2018lki}
H.~Mutuk, Adv. High Energy Phys. 2018 (2018) 8095653.

\bibitem{BaBar:2003oey}
B.~Aubert, et~al., Phys. Rev. Lett. 90 (2003) 242001.

\bibitem{CLEO:2003ggt}
D.~Besson, et~al., Phys. Rev. D 68 (2003) 032002, [Erratum: Phys.Rev.D 75, 119908 (2007)].

\bibitem{Belle:2003guh}
P.~Krokovny, et~al., Phys. Rev. Lett. 91 (2003) 262002.

\bibitem{BaBar:2003cdx}
B.~Aubert, et~al., Phys. Rev. D 69 (2004) 031101.

\bibitem{DiPierro:2001dwf}
M.~Di~Pierro, E.~Eichten, Phys. Rev. D 64 (2001) 114004.

\bibitem{PANDA:2021ozp}
G.~Barucca, et~al., Eur. Phys. J. A 57~(6) (2021) 184.

\bibitem{LHCb:2024nlg}
R.~Aaij, et~al., JHEP 04 (2024) 151.

\bibitem{Capriotti:2019huu}
L.~Capriotti, J. Phys. Conf. Ser. 1137~(1) (2019) 012004.

\bibitem{BESIII:2023wsc}
M.~Ablikim, et~al., Phys. Rev. Lett. 131~(15) (2023) 151903.

\bibitem{CMS:2019uhm}
A.~M. Sirunyan, et~al., Phys. Rev. Lett. 122~(13) (2019) 132001.

\bibitem{Jain:1993qh}
P.~Jain, H.~J. Munczek, Phys. Rev. D 48 (1993) 5403--5411.

\bibitem{Alkofer:2002bp}
R.~Alkofer, P.~Watson, H.~Weigel, Phys. Rev. D 65 (2002) 094026.

\bibitem{Bender:2002as}
A.~Bender, W.~Detmold, C.~Roberts, A.~W. Thomas, Phys. Rev. C65 (2002) 065203.

\bibitem{Chang:2009zb}
L.~Chang, C.~D. Roberts, Phys. Rev. Lett. 103 (2009) 081601.

\bibitem{Roberts:1994dr}
C.~D. Roberts, A.~G. Williams, Prog. Part. Nucl. Phys. 33 (1994) 477--575.

\bibitem{Savkli:1997kz}
C.~Savkli, F.~Tabakin, Nucl. Phys. A 628 (1998) 645--668.

\bibitem{Maris:2003vk}
P.~Maris, C.~D. Roberts, Int. J. Mod. Phys. E12 (2003) 297--365.

\bibitem{Watson:2004kd}
P.~Watson, W.~Cassing, P.~C. Tandy, Few Body Syst. 35 (2004) 129--153.

\bibitem{Eichmann:2008ae}
G.~Eichmann, R.~Alkofer, I.~C. Cloet, A.~Krassnigg, C.~D. Roberts, Phys. Rev. C 77 (2008) 042202.

\bibitem{Qin:2011dd}
S.-x. Qin, L.~Chang, Y.-x. Liu, C.~D. Roberts, D.~J. Wilson, Phys. Rev. C 84 (2011) 042202.

\bibitem{Fischer:2008wy}
C.~S. Fischer, R.~Williams, Phys. Rev. D 78 (2008) 074006.

\bibitem{Chang:2011ei}
L.~Chang, C.~D. Roberts, Phys. Rev. C 85 (2012) 052201.

\bibitem{Roberts:2011cf}
H.~L.~L. Roberts, L.~Chang, I.~C. Cloet, C.~D. Roberts, Few Body Syst. 51 (2011) 1--25.

\bibitem{Bashir:2012fs}
A.~Bashir, L.~Chang, I.~C. Cloet, B.~El-Bennich, Y.-X. Liu, et~al., Commun. Theor. Phys. 58 (2012) 79--134.

\bibitem{Eichmann:2013afa}
G.~Eichmann, J. Phys. Conf. Ser. 426 (2013) 012014.

\bibitem{Heupel:2014ina}
W.~Heupel, T.~Goecke, C.~S. Fischer, Eur. Phys. J. A50 (2014) 85.

\bibitem{Binosi:2014aea}
D.~Binosi, L.~Chang, J.~Papavassiliou, C.~D. Roberts, Phys. Lett. B742 (2015) 183--188.

\bibitem{Sanchis-Alepuz:2015tha}
H.~Sanchis-Alepuz, R.~Williams, J. Phys. Conf. Ser. 631~(1) (2015) 012064.

\bibitem{Williams:2015cvx}
R.~Williams, C.~S. Fischer, W.~Heupel, Phys. Rev. D93~(3) (2016) 034026.

\bibitem{Bedolla:2015mpa}
M.~A. Bedolla, J.~J. Cobos-Mart\'\i{}nez, A.~Bashir, Phys. Rev. D 92~(5) (2015) 054031.

\bibitem{Sanchis-Alepuz:2015qra}
H.~Sanchis-Alepuz, R.~Williams, Phys. Lett. B749 (2015) 592--596.

\bibitem{Li:2016mah}
B.-L. Li, L.~Chang, M.~Ding, C.~D. Roberts, H.-S. Zong, Phys. Rev. D 94~(9) (2016) 094014.

\bibitem{Binosi:2016rxz}
D.~Binosi, L.~Chang, J.~Papavassiliou, S.-X. Qin, C.~D. Roberts, Phys. Rev. D93~(9) (2016) 096010.

\bibitem{Binosi:2016xxu}
D.~Binosi, C.~D. Roberts, J.~Rodriguez-Quintero, Phys. Rev. D 95~(11) (2017) 114009.

\bibitem{Serna:2017nlr}
F.~E. Serna, B.~El-Bennich, G.~a. Krein, Phys. Rev. D 96~(1) (2017) 014013.

\bibitem{Miramontes:2019mco}
A.~S. Miramontes, H.~Sanchis-Alepuz, Eur. Phys. J. A 55~(10) (2019) 170.

\bibitem{Eichmann:2020oqt}
G.~Eichmann, C.~S. Fischer, W.~Heupel, N.~Santowsky, P.~C. Wallbott, Few Body Syst. 61~(4) (2020) 38.

\bibitem{Wallbott:2020jzh}
P.~C. Wallbott, G.~Eichmann, C.~S. Fischer, Phys. Rev. D 102~(5) (2020) 051501.

\bibitem{Miramontes:2021xgn}
A.~S. Miramontes, H.~Sanchis~Alepuz, R.~Alkofer, Phys. Rev. D 103~(11) (2021) 116006.

\bibitem{Gutierrez-Guerrero:2021rsx}
L.~X. Guti\'errez-Guerrero, G.~Paredes-Torres, A.~Bashir, Phys. Rev. D 104~(9) (2021) 094013.

\bibitem{Yin:2019bxe}
P.-L. Yin, C.~Chen, G.~a. Krein, C.~D. Roberts, J.~Segovia, S.-S. Xu, Phys. Rev. D 100~(3) (2019) 034008.

\bibitem{Eichmann:2023tjk}
G.~Eichmann, A.~G\'omez, J.~Horak, J.~M. Pawlowski, J.~Wessely, N.~Wink, Phys. Rev. D 109~(9) (2024) 096024.

\bibitem{Raya:2024ejx}
K.~Raya, A.~Bashir, D.~Binosi, C.~D. Roberts, J.~Rodr\'\i{}guez-Quintero, Few Body Syst. 65~(2) (2024) 60.

\bibitem{Marciano:1977su}
W.~J. Marciano, H.~Pagels, Phys. Rept. 36 (1978) 137.

\bibitem{Ball:1980ay}
J.~S. Ball, T.-W. Chiu, Phys. Rev. D22 (1980) 2542.

\bibitem{vonSmekal:1997ohs}
L.~von Smekal, R.~Alkofer, A.~Hauck, Phys. Rev. Lett. 79 (1997) 3591--3594.

\bibitem{Boucaud:2008gn}
P.~Boucaud, F.~De~Soto, J.~Leroy, A.~Le~Yaouanc, J.~Micheli, et~al., Phys. Rev. D79 (2009) 014508.

\bibitem{vonSmekal:2009ae}
L.~von Smekal, K.~Maltman, A.~Sternbeck, Phys. Lett. B 681 (2009) 336--342.

\bibitem{Blossier:2012ef}
B.~Blossier, P.~Boucaud, M.~Brinet, F.~De~Soto, X.~Du, V.~Morenas, O.~Pene, K.~Petrov, J.~Rodriguez-Quintero, Phys. Rev. Lett. 108 (2012) 262002.

\bibitem{Zafeiropoulos:2019flq}
S.~Zafeiropoulos, P.~Boucaud, F.~De~Soto, J.~Rodr\'{\i}guez-Quintero, J.~Segovia, Phys. Rev. Lett. 122~(16) (2019) 162002.

\bibitem{Maris:1997hd}
P.~Maris, C.~D. Roberts, P.~C. Tandy, Phys. Lett. B420 (1998) 267--273.

\bibitem{Aguilar:2021okw}
A.~C. Aguilar, C.~O. Ambr\'osio, F.~De~Soto, M.~N. Ferreira, B.~M. Oliveira, J.~Papavassiliou, J.~Rodr\'\i{}guez-Quintero, Phys. Rev. D 104~(5) (2021) 054028.

\bibitem{Mitter:2014wpa}
M.~Mitter, J.~M. Pawlowski, N.~Strodthoff, Phys. Rev. D91 (2015) 054035.

\bibitem{Cyrol:2016tym}
A.~K. Cyrol, L.~Fister, M.~Mitter, J.~M. Pawlowski, N.~Strodthoff, Phys. Rev. D94~(5) (2016) 054005.

\bibitem{Cyrol:2017ewj}
A.~K. Cyrol, M.~Mitter, J.~M. Pawlowski, N.~Strodthoff, Phys. Rev. D97~(5) (2018) 054006.

\bibitem{Huber:2018ned}
M.~Q. Huber, Phys. Rept. 879 (2020) 1--92.

\bibitem{Gao:2021wun}
F.~Gao, J.~Papavassiliou, J.~M. Pawlowski, Phys. Rev. D 103~(9) (2021) 094013.

\bibitem{Ferreira:2023fva}
M.~N. Ferreira, J.~Papavassiliou, Particles 6~(1) (2023) 312--363.

\bibitem{Binosi:2009qm}
D.~Binosi, J.~Papavassiliou, Phys. Rept. 479 (2009) 1--152.

\bibitem{Aguilar:2009nf}
A.~C. Aguilar, D.~Binosi, J.~Papavassiliou, J.~Rodriguez-Quintero, Phys. Rev. D80 (2009) 085018.

\bibitem{Binosi:2016nme}
D.~Binosi, C.~Mezrag, J.~Papavassiliou, C.~D. Roberts, J.~Rodriguez-Quintero, Phys. Rev. D96~(5) (2017) 054026.

\bibitem{Bjorken:1965zz}
J.~D. Bjorken, S.~D. Drell, {Relativistic quantum fields}, International Series In Pure and Applied Physics, McGraw-Hill, New York, 1965.

\bibitem{Baym:1961zz}
G.~Baym, L.~P. Kadanoff, Phys. Rev. 124 (1961) 287--299.

\bibitem{Cornwall:1973ts}
J.~Cornwall, R.~Norton, Phys. Rev. D 8 (1973) 3338--3346.

\bibitem{Cornwall:1974vz}
J.~M. Cornwall, R.~Jackiw, E.~Tomboulis, Phys. Rev. D 10 (1974) 2428--2445.

\bibitem{Berges:2004pu}
J.~Berges, Phys. Rev. D 70 (2004) 105010.

\bibitem{Alkofer:2008tt}
R.~Alkofer, C.~S. Fischer, F.~J. Llanes-Estrada, K.~Schwenzer, Annals Phys. 324 (2009) 106--172.

\bibitem{Carrington:2010qq}
M.~E. Carrington, Y.~Guo, Phys. Rev. D 83 (2011) 016006.

\bibitem{York:2012ib}
M.~C.~A. York, G.~D. Moore, M.~Tassler, JHEP 06 (2012) 077.

\bibitem{Pawlowski:2005xe}
J.~M. Pawlowski, Annals Phys. 322 (2007) 2831--2915.

\bibitem{Davydychev:2000rt}
A.~I. Davydychev, P.~Osland, L.~Saks, Phys. Rev. D63 (2001) 014022.

\bibitem{Aguilar:2018csq}
A.~C. Aguilar, M.~N. Ferreira, C.~T. Figueiredo, J.~Papavassiliou, Phys. Rev. D99 (2019) 034026.

\bibitem{Aguilar:2010cn}
A.~C. Aguilar, J.~Papavassiliou, Phys. Rev. D83 (2011) 014013.

\bibitem{Aguilar:2016lbe}
A.~C. Aguilar, J.~C. Cardona, M.~N. Ferreira, J.~Papavassiliou, Phys. Rev. D96~(1) (2017) 014029.

\bibitem{Taylor:1971ff}
J.~Taylor, Nucl. Phys. B 33 (1971) 436--444.

\bibitem{Qin:2014vya}
S.-X. Qin, C.~D. Roberts, S.~M. Schmidt, Phys. Lett. B 733 (2014) 202--208.

\bibitem{Roberts:2015lja}
C.~D. Roberts, J. Phys. Conf. Ser. 706~(2) (2016) 022003.

\bibitem{Xu:2022kng}
Z.-N. Xu, Z.-Q. Yao, S.-X. Qin, Z.-F. Cui, C.~D. Roberts, Eur. Phys. J. A 59~(3) (2023) 39.

\bibitem{Gell-Mann:1968hlm}
M.~Gell-Mann, R.~J. Oakes, B.~Renner, Phys. Rev. 175 (1968) 2195--2199.

\bibitem{Maris:1999nt}
P.~Maris, P.~C. Tandy, Phys. Rev. C60 (1999) 055214.

\bibitem{Eichmann:2016yit}
G.~Eichmann, H.~Sanchis-Alepuz, R.~Williams, R.~Alkofer, C.~S. Fischer, Prog. Part. Nucl. Phys. 91 (2016) 1--100.

\bibitem{Sanchis-Alepuz:2017jjd}
H.~Sanchis-Alepuz, R.~Williams, Comput. Phys. Commun. 232 (2018) 1--21.

\bibitem{Fischer:2005en}
C.~Fischer, P.~Watson, W.~Cassing, Phys. Rev. D72 (2005) 094025.

\bibitem{Krassnigg:2008bob}
A.~Krassnigg, PoS CONFINEMENT8 (2008) 075.

\bibitem{ParticleDataGroup:2022pth}
R.~L. Workman, et~al., PTEP 2022 (2022) 083C01.

\bibitem{Lubicz:2017asp}
V.~Lubicz, A.~Melis, S.~Simula, Phys. Rev. D 96~(3) (2017) 034524.

\bibitem{Dowdall:2012ab}
R.~J. Dowdall, C.~T.~H. Davies, T.~C. Hammant, R.~R. Horgan, Phys. Rev. D 86 (2012) 094510.

\bibitem{Cichy:2016bci}
K.~Cichy, M.~Kalinowski, M.~Wagner, Phys. Rev. D 94~(9) (2016) 094503.

\bibitem{Mathur:2018epb}
N.~Mathur, M.~Padmanath, S.~Mondal, Phys. Rev. Lett. 121~(20) (2018) 202002.

\bibitem{Donald:2012ga}
G.~C. Donald, C.~T.~H. Davies, R.~J. Dowdall, E.~Follana, K.~Hornbostel, J.~Koponen, G.~P. Lepage, C.~McNeile, Phys. Rev. D 86 (2012) 094501.

\bibitem{Bazavov:2017lyh}
A.~Bazavov, et~al., Phys. Rev. D 98~(7) (2018) 074512.

\bibitem{Davies:2010ip}
C.~T.~H. Davies, C.~McNeile, E.~Follana, G.~P. Lepage, H.~Na, J.~Shigemitsu, Phys. Rev. D 82 (2010) 114504.

\bibitem{Hughes:2017spc}
C.~Hughes, C.~T.~H. Davies, C.~J. Monahan, Phys. Rev. D 97~(5) (2018) 054509.

\bibitem{McNeile:2012qf}
C.~McNeile, C.~T.~H. Davies, E.~Follana, K.~Hornbostel, G.~P. Lepage, Phys. Rev. D 86 (2012) 074503.

\bibitem{Boyle:2017jwu}
P.~A. Boyle, L.~Del~Debbio, A.~Jüttner, A.~Khamseh, F.~Sanfilippo, J.~T. Tsang, J. High Energy Phys. 12 (2017) 008.

\bibitem{Christ:2014uea}
N.~H. Christ, J.~M. Flynn, T.~Izubuchi, T.~Kawanai, C.~Lehner, A.~Soni, R.~S. Van~de Water, O.~Witzel, Phys. Rev. D 91~(5) (2015) 054502.

\bibitem{Rojas:2014aka}
E.~Rojas, B.~El-Bennich, J.~P. B.~C. de~Melo, Phys. Rev. D 90 (2014) 074025.

\bibitem{Mojica:2017tvh}
F.~F. Mojica, C.~E. Vera, E.~Rojas, B.~El-Bennich, Phys. Rev. D 96~(1) (2017) 014012.

\bibitem{Qin:2020jig}
S.-X. Qin, C.~D. Roberts, Chin. Phys. Lett. 38~(7) (2021) 071201.

\end{thebibliography}

%






\end{document}